\documentclass[12pt]{article}%
\usepackage{amsmath,amssymb,amsthm}
\begin{document}

\title{Metric Tensor Vs. Metric Extensor\thanks{published: \textit{Advances in
Applied Clifford Algebras} \textbf{11}(S3), 41-48 (2001).}}
\author{V. V. Fern\'{a}ndez$^{1}\thanks{e-mail: vvf@ime.unicamp.br}$, A. M.
Moya$^1\thanks{e-mail: moya@ime.unicamp.br}$ and W. A. Rodrigues
Jr.$^1,2\thanks{e-mail: walrod@ime.unicamp.br or walrod@mpc.com.br }$\\$\hspace{-0.5cm}^{1}$ Institute of Mathematics, Statistics and Scientific Computation\\IMECC-UNICAMP CP 6065\\13083-970 Campinas-SP, Brazil\\$^{2}$ Department of Mathematical Sciences, University of Liverpool\\Liverpool, L69 3BX, UK}
\date{10/30/2001}
\maketitle

\begin{abstract}
In this paper we give a comparison between the formulation of the concept of
metric for a real vector space of finite dimension in terms of \emph{tensors}
and \emph{extensors}. A nice property of metric extensors is that they have
inverses which are also themselves metric extensors. This property is not
shared by metric tensors because tensors do \emph{not} have inverses. We
relate the definition of determinant of a metric extensor with the classical
determinant of the corresponding matrix associated to the metric tensor in a
given vector basis. Previous identifications of these concepts are
equivocated. The use of metric extensor permits sophisticated calculations
without the introduction of matrix representations.

\end{abstract}
\tableofcontents

\section{Introduction}

This is the third paper of a series of seven. Here, we explore the concept of
metric on a $n$-dimensional real vector space $V$ by using the concepts of
tensors and extensors. We show that for each metric tensor $G$, there is an
unique metric extensor $g,$ and vice versa. The metric extensor is a
fundamental tool for future developments that we have in mind, and which are
going to be presented in following papers of this series and subsequent series
of papers. Besides this fact, it is worth to emphasize here that the concept
of metric extensor has a prior status in the foundations of linear algebra
relative to its corresponding metric tensor. This is because a metric extensor
$g$ has an inverse $g^{-1}$ which is of course itself a metric extensor. As a
simple application of this concept, we use $g$ and $g^{-1},$ to recall some
results involving the well-known metric isomorphism between $V$ and $V^{*}$
induced by $G.$

\section{Standard Isomorphism}

Let $\{b_{k}\}$ be an arbitrary, but fixed, basis for $V$ and $\{\beta^{k}\} $
its corresponding dual basis for $V^{*},$ i.e., $\beta^{k}(b_{j})=\delta
_{j}^{k}.$ There exists a linear isomorphism between $V$ and $V^{*}$ realized
by the linear mappings $V\ni v\mapsto\underset{b}{\iota}(v)\in V^{*}$ and
$V^{*}\ni\omega\mapsto\underset{\beta}{\iota^{-1}}(\omega)\in V$ such that
\begin{align}
\underset{b}{\iota}(v)  & =\overset{n}{\underset{k=1}{\sum}}\beta^{k}%
(v)\beta^{k},\label{3.1a}\\
\underset{\beta}{\iota^{-1}}(\omega)  & =\overset{n}{\underset{k=1}{\sum}%
}\omega(b_{k})b_{k}.\label{3.1b}%
\end{align}
A suggested by the notations, $\underset{\beta}{\iota^{-1}}$ is inverse
mapping of $\underset{b}{\iota}.$%

\begin{proof}
The linearity property for both $\underset{b}{\iota}$ and $\underset{\beta
}{\iota^{-1}}$ holds. We must prove that $\underset{\beta}{\iota^{-1}}%
\circ\underset{b}{\iota}=i_{V}$ and $\underset{b}{\iota}\circ\underset{\beta
}{\iota^{-1}}=i_{V^{*}},$ where $i_{V}$ and $i_{V^{*}}$ are the identity
mappings in $V$ and $V^{*},$ respectively.

First take $v\in V,$ using the linearity property of $\underset{\beta}%
{\iota^{-1}},$ eq.(\ref{3.1b}), the duality condition $\beta^{k}(b_{j}%
)=\delta_{j}^{k}$ and the elementary expansion for vectors, i.e., $v=\beta
^{k}(v)b_{k,}$ we have
\[
\underset{\beta}{\iota^{-1}}\circ\underset{b}{\iota}(v)=\overset{n}%
{\underset{k=1}{\sum}}\beta^{k}(v)\iota^{-1}(\beta^{k})=\overset{n}%
{\underset{k=1}{\sum}}\beta^{k}(v)\overset{n}{\underset{s=1}{\sum}}\delta
_{s}^{k}b_{s}=\beta^{k}(v)b_{k}=v,
\]
hence, $\underset{\beta}{\iota^{-1}}\circ\underset{b}{\iota}=i_{V}$.

Now take $\omega\in V^{*},$ using the linearity property of $\underset
{b}{\iota},$ eq.(\ref{3.1a}), the duality condition $\beta^{k}(b_{j}%
)=\delta_{j}^{k}$ and the elementary expansion for forms, i.e., $\omega
=\omega(b_{k})\beta^{k},$ we have
\[
\underset{b}{\iota}\circ\underset{\beta}{\iota^{-1}}(\omega)=\overset
{n}{\underset{k=1}{\sum}}\omega(b_{k})\iota(b_{k})=\overset{n}{\underset
{k=1}{\sum}}\omega(b_{k})\overset{n}{\underset{s=1}{\sum}}\delta_{k}^{s}%
\beta^{s}=\omega(b_{k})\beta^{k}=\omega,
\]
hence, $\underset{b}{\iota}\circ\underset{\beta}{\iota^{-1}}=i_{V^{*}}.$
\end{proof}

This linear isomorphism between $V$ and $V^{*}$ induced by the pair of bases
$(\{b_{k}\},\{\beta^{k}\})$ will be called a\emph{\ standard} isomorphism.

As introduced above this linear isomorphism is the unique satisfying the
following two conditions: $\underset{b}{\iota}(b_{k})=\beta^{k}$ and
$\underset{\beta}{\iota^{-1}}(\beta^{k})=b_{k}.$

\subsection{$b$-Scalar Products}

We recall from \cite{1} (paper I on this series) that we can define a
$b$-scalar product of vectors $v,$ $w\in V$ by
\begin{equation}
v\underset{b}{\cdot}w=\underset{b}{\iota}(v)(w)=\overset{n}{\underset
{k=1}{\sum}}\beta^{k}(v)\beta^{k}(w).\label{3.1.1}%
\end{equation}

Also, a $b$-scalar product of forms $\omega$ and $\sigma$ can be defined by
\begin{equation}
\omega\underset{\beta}{\cdot}\sigma=\sigma(\underset{\beta}{\iota^{-1}}%
(\omega))=\overset{n}{\underset{k=1}{\sum}}\omega(b_{k})\sigma(b_{k}%
).\label{3.1.2}%
\end{equation}

The products defined by eq.(\ref{3.1.1}) and eq.(\ref{3.1.2}) are well-defined
scalar products in $V$ and $V^{*},$ associated with the arbitrary pair of
bases $(\{b_{k}\},\{\beta^{k}\})$, and also these scalar products are positive
definite, e.g., for the $b$-scalar product of vectors: $v\underset{b}{\cdot
}v\geq0$ for all $v,$ and $v\underset{b}{\cdot}v=0 $ if and only if $v=0.$

The vector basis $\{b_{k}\}$ and the form basis $\{\beta^{k}\}$ are both
orthonormal in each of spaces $V$ and $V^{*},$ respectively, i.e.,
\begin{align}
b_{j}\underset{b}{\cdot}b_{k}  & =\delta_{jk},\label{3.1.2a}\\
\beta^{j}\underset{b}{\cdot}\beta^{k}  & =\delta^{jk}.\label{3.1.2b}%
\end{align}

\begin{proof}

The use of eq.(\ref{3.1.1}) and the duality condition of $(\{b_{k}%
\},\{\beta^{k}\})$ gives
\[
b_{j}\underset{b}{\cdot}b_{k}=\overset{n}{\underset{p=1}{\sum}}\beta^{p}%
(b_{j})\beta^{p}(b_{k})=\overset{n}{\underset{p=1}{\sum}}\delta_{j}^{p}%
\delta_{k}^{p}=\delta_{jk},
\]
by using eq.(\ref{3.1.2}) and once again the duality condition of
$(\{b_{k}\},\{\beta^{k}\})$ we get
\[
\beta^{j}\underset{b}{\cdot}\beta^{k}=\overset{n}{\underset{q=1}{\sum}}%
\beta^{j}(b_{q})\beta^{k}(b_{q})=\overset{n}{\underset{q=1}{\sum}}\delta
_{q}^{j}\delta_{q}^{k}=\delta^{jk}.
\]
\end{proof}

In what follows we will use the simplified notation $(\cdot)$ for the
$b$-scalar product $(\underset{b}{\cdot}).$

\subsection{$b$-Reciprocal Bases}

Let $(\{e_{k}\},\{\varepsilon^{k}\})$ be any bases for $V$ and $V^{*}$, the
second one being dual basis of the first one, i.e., $\varepsilon^{k}%
(e_{j})=\delta_{j}^{k}.$ Associated to it we can introduce another pair of
bases $(\{e^{k}\},\{\varepsilon_{k}\}),$ the first one for $V$ and the second
one for $V^{*},$ given by
\begin{align}
e^{k}  & =\iota^{-1}(\varepsilon^{k})=\overset{n}{\underset{s=1}{\sum}%
}\varepsilon^{k}(b_{s})b_{s},\label{3.1.3}\\
\varepsilon_{k}  & =\iota(e_{k})=\overset{n}{\underset{s=1}{\sum}}\beta
^{s}(e_{k})\beta^{s}.\label{3.1.4}%
\end{align}

Since $\iota$ and $\iota^{-1}$ are linear mappings between $V$ and $V^{*},$
the $n$ vectors $e^{1},\ldots,e^{n}\in V$ and the $n$ forms $\varepsilon
_{1},\ldots,\varepsilon_{n}\in V^{*}$ are linearly independent, hence they are
a vector basis for $V$ and a form basis for $V^{*},$ respectively.

\textbf{i.} The pairs of bases $(\{e_{k}\},\{e^{k}\})$ and $(\{\varepsilon
^{k}\},\{\varepsilon_{k}\})$ satisfy the remarkable $b$-scalar product
conditions
\begin{align}
e_{k}\cdot e^{l}  & =\delta_{k}^{l},\label{3.1.5}\\
\varepsilon^{k}\cdot\varepsilon_{l}  & =\delta_{l}^{k}.\label{3.1.6}%
\end{align}
\begin{proof}
By straightforward calculation, employing eq.(\ref{3.1.1}) and eq.(\ref{3.1.3}%
), and the duality condition of $(\{e_{k}\},\{\varepsilon^{k}\})$ we have
\begin{align*}
e_{k}\cdot e^{l}  & =e^{l}\cdot e_{k}\\
& =\iota(e^{l})(e_{k})=\iota\circ\iota^{-1}(\varepsilon^{l})(e_{k}%
)=\varepsilon^{l}(e_{k})=\delta_{k}^{l},
\end{align*}
and employing eq.(\ref{3.1.2}) and eq.(\ref{3.1.4}), and once again the
duality condition of $(\{e_{k}\},\{\varepsilon^{k}\})$ we get
\[
\varepsilon_{k}\cdot\varepsilon^{l}=\varepsilon^{l}(\iota^{-1}(\varepsilon
_{k}))=\varepsilon^{l}(\iota^{-1}\circ\iota(e_{k}))=\varepsilon^{l}%
(e_{k})=\delta_{k}^{l}.
\]
\end{proof}

Due to the property expressed by eq.(\ref{3.1.5}) the vector bases $\{e_{k}\}
$ and $\{e^{k}\}$ are called a $b$-\emph{reciprocal bases} of $V.$ The form
bases $\{\varepsilon^{k}\}$ and $\{\varepsilon_{k}\}$ because the property
given by eq.(\ref{3.1.6}) are called a $b$-\emph{reciprocal bases} of $V^{*}.
$

\textbf{ii.} $\{\varepsilon_{k}\}$ is dual basis of $\{e^{k}\},$ i.e.,
\begin{equation}
\varepsilon_{k}(e^{j})=\delta_{k}^{j}.\label{3.1.7}%
\end{equation}
\begin{proof}
The equations (\ref{3.1.4}), (\ref{3.1.1}) and (\ref{3.1.5}) yield
\[
\varepsilon_{k}(e^{j})=\iota(e_{k})(e^{j})=e_{k}\cdot e^{j}=\delta_{k}^{j},
\]
Also, the equations (\ref{3.1.3}), (\ref{3.1.2}) and (\ref{3.1.6}) yield
\[
\varepsilon_{k}(e^{j})=\varepsilon_{k}(\iota^{-1}(\varepsilon^{j}%
))=\varepsilon^{j}\cdot\varepsilon_{k}=\delta_{k}^{j}.
\]
\end{proof}

\section{$2$-Tensors vs. $(1,1)$-Extensors}

For any $2$-tensor $T\in T_{2}(V),$ there exists an unique $(1,1)$-extensor
$t\in ext_{1}^{1}(V)$, such that for all vectors $v,w\in V$%
\begin{equation}
T(v,w)=t(v)\cdot w.\label{3.2}%
\end{equation}
\begin{proof}
We will prove that $t(v)=T(v,e_{k})e^{k}$ (or, also $t(v)=T(v,e^{k})e_{k}$),
where $(\{e_{k}\},\{e^{k}\})$ is an arbitrary pair of $b$-reciprocal bases for
$V,$ satisfies eq.(\ref{3.2}). Indeed, by using an expansion formula for
vectors \cite{1} and the linearity property of tensors, we have
\[
t(v)\cdot w=T(v,e_{k})e^{k}\cdot w=T(v,e^{k}\cdot we_{k})=T(v,w).
\]

Now, $t$ is unique. Indeed suppose that there is another $t^{\prime}\in
ext_{1}^{1}(V)$ for which $T(v,w)=t^{\prime}(v)\cdot w$ for arbitrary $v$ and
$w$. Then, using the expansion formula for vectors it follows that
\[
t^{\prime}(v)=t^{\prime}(v)\cdot e_{k}e^{k}=T(v,e_{k})e^{k}=t(v).
\]

Thus, the \emph{existence} and \emph{uniqueness }of such a $(1,1)$-extensor
satisfying eq.(\ref{3.2}) is established$.$
\end{proof}
The above theorem means that there exists an unique linear isomorphism between
the vector spaces $T_{2}(V)$ and $ext_{1}^{1}(V)$ such that the $jk$-th
covariant components of $T$ with respect to $\{e_{k}\}$ equals the $jk$-matrix
element of $t$ with respect to $\{e_{k}\},$ i.e., $T(e_{j},e_{k}%
)=t(e_{j})\cdot e_{k}.$

$T$ is \emph{symmetric} (or, \emph{skew-symmetric}) if and only if $t$ is
\emph{adjoint symmetric} (or, \emph{adjoint skew-symmetric}), i.e.,
\begin{equation}
T(v,w)=T(w,v),\text{ for all }v,w\in V\Leftrightarrow t=t^{\dagger
},\label{3.2.0a}%
\end{equation}
or,
\begin{equation}
T(v,w)=-T(w,v),\text{ for all }v,w\in V\Leftrightarrow t=-t^{\dagger
}.\label{3.2.0b}%
\end{equation}

Let $T_{jk}$ be the $jk$-entries of the $n\times n$ real matrix associated to
$T$ with respect to a basis $\{e_{k}\},$ i.e., $T_{jk}=T(e_{j},e_{k}).$ The
classical determinant $\det\left[  T_{jk}\right]  $ and $\det[t]$ are related
by the remarkable formula\footnote{Note that some previous papers that
appeared in the literature and which use the concept of extensor \cite{3} have
made a wrong identification between the classical determinant of the matrix
formed with the elements $T_{jk},$ and $\det[t]$.}
\begin{equation}
\det\left[  T_{jk}\right]  =\det[t](e_{1}\wedge\ldots\wedge e_{n})\cdot
(e_{1}\wedge\ldots\wedge e_{n}).\label{3.2.1}%
\end{equation}
\begin{proof}
By definition of classical determinant of $n\times n$ real matrix. Using
eq.(\ref{3.2}), the formula $(v_{1}\wedge\ldots v_{k})\cdot(w_{1}\wedge\ldots
w_{k})=\epsilon^{s_{1}\ldots s_{k}}v_{1}\cdot w_{s_{1}}\ldots v_{k}\cdot
w_{s_{k}}$ and the property $\underline{t}(v_{1}\wedge\ldots\wedge
v_{k})=t(v_{1})\wedge\ldots t(v_{k}),$ where $v_{1},\ldots,v_{k}$ and
$w_{1},\ldots,w_{k}$ are vectors. We have
\begin{align*}
\left|  T_{jk}\right|   & =\epsilon^{s_{1}\ldots s_{n}}T_{1s_{1}}\ldots
T_{ns_{n}}=\epsilon^{s_{1}\ldots s_{n}}T(e_{1},e_{s_{1}})\ldots T(e_{n}%
,e_{s_{n}})\\
& =\epsilon^{s_{1}\ldots s_{n}}t(e_{1})\cdot e_{s_{1}}\ldots t(e_{n})\cdot
e_{s_{n}}=(t(e_{1})\wedge\ldots t(e_{n}))\cdot(e_{1}\wedge\ldots e_{n})\\
& =\underline{t}(e_{1}\wedge\ldots e_{n})\cdot(e_{1}\wedge\ldots e_{n}),
\end{align*}
hence, by definition of $\det[t],$ i.e., $\underline{t}(I)=\det[t]I$ for all
non-null pseudoscalar $I,$ the required result follows.
\end{proof}

The $n\times n$ real matrix $T_{jk}$ has inverse (i.e., there exists an unique
$n\times n$ real matrix whose $jk$-entries, say $T^{jk},$ satisfying
$T^{js}T_{sk}=\delta_{k}^{j}$ and $T_{js}T^{sk}=\delta_{j}^{k}$) if and only
if $t$ has inverse, say $t^{-1}$ (i.e., $t^{-1}$ is the unique $(1,1)$%
-extensor that satisfies $t^{-1}\circ t=t\circ t^{-1}=i_{V}$). The real
numbers $T^{jk}$ are equal to the $jk$-matrix elements of $t^{-1},$ i.e.,
\begin{equation}
T^{jk}=t^{-1}(e^{j})\cdot e^{k}.\label{3.2.2}%
\end{equation}
\begin{proof}
The first statement is an immediate consequence of eq.(\ref{3.2.1}). In order
to prove the second statement we shall use the expansion formula for $v\in
V:v=v\cdot e^{s}e_{s},$ to get
\begin{align*}
T^{js}T_{sk}  & =t^{-1}(e^{j})\cdot e^{s}t(e_{s})\cdot e_{k}=t(t^{-1}%
(e^{j})\cdot e^{s}e_{s})\cdot e_{k}\\
& =t\circ t^{-1}(e^{j})\cdot e_{k}=e^{j}\cdot e_{k},
\end{align*}
hence, by the $b$-reciprocity condition of $(\{e_{k}\},\{e^{k}\}),$ i.e.,
$e_{k}\cdot e^{l}=\delta_{k}^{l},$ we have that $T^{js}T_{sk}=\delta_{k}^{j}
$. Analogously it can be proved that $T_{js}T^{sk}=\delta_{j}^{k}.$
\end{proof}

\section{Metric Isomorphism}

We recall some definitions of the theory of ordinary linear algebra.

Let $G$ be a (covariant) metric tensor over $V,$ i.e., $G\in T_{2}(V)$ such
that: $G$ is symmetric ($G(v,w)=G(w,v)$) and $G$ is non-degenerate
($\det\left[  G_{jk}\right]  \neq0,$ $G_{jk}=G(e_{j},e_{k})$ where $\{e_{k}\}$
is any basis for $V$).

According to equations (\ref{3.2}), (\ref{3.2.0a}) and (\ref{3.2.1}) there
exists a $(1,1)$-extensor over $V,$ say $g,$ symmetric ($g=g^{\dagger}$) and
non-degenerate ($\det[g]\neq0$), such that for all $v,w\in V:$
$G(v,w)=g(v)\cdot w$. By our previous nomenclature $g\in ext_{1}^{1}(V)$ is
called a \emph{metric extensor over }$V.$

This algebraic object \emph{codifies} all the information contained into the
classical concept of metric which appears in ordinary linear algebra.

The well-defined $2$-tensor over $V^{*},$ $G^{*}\in T^{2}(V)$ such that
$G^{*}(\omega,\sigma)=G^{jk}\omega(e_{j})\sigma(e_{k}),$ where $G^{jk}$ are
the $jk$-entries of the inverse matrix of $G_{jk}=G(e_{j},e_{k}),$ is said to
be the (contravariant) metric tensor over $V^{*}.$

There exists a linear isomorphism between $V$ and $V^{*}$ realized by the
linear mappings $V\ni v\mapsto\underset{G}{\iota}(v)\in V^{*}$and $V^{*}%
\ni\omega\mapsto\underset{G}{\iota}^{-1}(\omega)\in V$ such that
\begin{align}
\underset{G}{\iota}(v)(w)  & =G(v,w),\label{3.3a}\\
\sigma(\underset{G}{\iota}^{-1}(\omega))  & =G^{*}(\omega,\sigma).\label{3.3b}%
\end{align}
As the notations point out $\underset{G}{\iota}^{-1},$ is the inverse mapping
of $\underset{G}{\iota}.$%

\begin{proof}
The linearity property for both $\underset{G}{\iota}$ and $\underset{G}{\iota
}^{-1}$ holds. We must prove that $\underset{G}{\iota}^{-1}\circ\underset
{G}{\iota}=i_{V}$ and $\underset{G}{\iota}\circ\underset{G}{\iota}%
^{-1}=i_{V^{*}},$ where $i_{V}$ and $i_{V^{*}}$ are the identity functions in
$V$ and $V^{*},$ respectively.

Let $(\{e_{k}\},\{\varepsilon^{k}\})$ be an arbitrary pair of dual basis (of
$V$ and $V^{*}$), i.e., $\varepsilon^{k}(e_{j})=\delta_{j}^{k}.$ Recall the
elementary expansions for vectors and forms, i.e., $v=\varepsilon^{k}(v)e_{k}$
and $\omega=\omega(e_{k})\varepsilon^{k}.$

Take $v\in V$ and $\omega\in V^{*},$ $\underset{G}{\iota}(v)$ and
$\underset{G}{\iota}^{-1}(\omega)$. Some well-known formulas involving the
matrix elements of $G$ and $G^{*}$ can be written as follows
\begin{align}
\underset{G}{\iota}(v)  & =\underset{G}{\iota}(v)(e_{k})\varepsilon
^{k}=G(v,e_{k})\varepsilon^{k}\nonumber\\
& =G(\varepsilon^{j}(v)e_{j},e_{k})\varepsilon^{k}=G(e_{j},e_{k}%
)\varepsilon^{j}(v)\varepsilon^{k},\nonumber\\
\underset{G}{\iota}(v)  & =G_{jk}\varepsilon^{j}(v)\varepsilon^{k}%
.\label{3.first}%
\end{align}

Also,
\begin{align}
\underset{G}{\iota}^{-1}(\omega)  & =\varepsilon^{k}(\underset{G}{\iota}%
^{-1}(\omega))e_{k}=G^{*}(\omega,\varepsilon^{k})e_{k}\nonumber\\
& =G^{*}(\omega(e_{j})\varepsilon^{j},\varepsilon^{k})e_{k}=G^{*}%
(\varepsilon^{j},\varepsilon^{k})\omega(e_{j})e_{k}\nonumber\\
\underset{G}{\iota}^{-1}(\omega)  & =G^{jk}\omega(e_{j})e_{k}.\label{3.second}%
\end{align}

Next, take $v\in V$. Using eq.(\ref{3.first}), eq.(\ref{3.3a}) and the formula
for expansion of vectors we have
\begin{align*}
\underset{G}{\iota}^{-1}\circ\underset{G}{\iota}(v)  & =G^{jk}\underset
{G}{\iota}(v)(e_{j})e_{k}=G^{jk}G(v,e_{j})e_{k}=G^{jk}G(e_{s},e_{j}%
)\varepsilon^{s}(v)e_{k}\\
& =G^{jk}G_{sj}\varepsilon^{s}(v)e_{k}=\delta_{s}^{k}\varepsilon^{s}%
(v)e_{k}=\varepsilon^{k}(v)e_{k}=v,
\end{align*}
hence, $\underset{G}{\iota}^{-1}\circ\underset{G}{\iota}=i_{V}.$

By the same way, taking an arbitrary $\omega\in V^{*}$ and employing
eq.(\ref{3.second}), eq.(\ref{3.3b}) and the elementary expansion for forms,
we finally get $\underset{G}{\iota}\circ\underset{G}{\iota}^{-1}=i_{V^{*}}.$
\end{proof}

The linear isomorphism between $V$ and $V^{*}$ showed above, which is induced
by the (covariant) metric tensor $G$ is usually called a \emph{metric
isomorphism}.

The inverse mappings $\underset{G}{\iota}$ and $\underset{G}{\iota}^{-1}$ can
be written in suggestive forms involving the $(1,1)$-extensors $g$ and
$g^{-1},$ and the inverse mappings $\iota$ and $\iota^{-1}$ of the standard
isomorphism, i.e.,
\begin{align}
\underset{G}{\iota}  & =\iota\circ g,\label{3.3.1}\\
\underset{G}{\iota}^{-1}  & =g^{-1}\circ\iota^{-1}.\label{3.3.2}%
\end{align}
\begin{proof}
First take $v\in V,$ using equations (\ref{3.3a}), (\ref{3.2}) and
(\ref{3.1.3}) we have
\begin{align*}
\underset{G}{\iota}(v)  & =\underset{G}{\iota}(v)(e_{k})\varepsilon
^{k}=G(v,e_{k})\varepsilon^{k}=g(v)\cdot e_{k}\varepsilon^{k}\\
& =g(v)\cdot e_{k}\iota(e^{k})=\iota(g(v)\cdot e_{k}e^{k})\\
& =\iota\circ g(v),
\end{align*}
hence, $\underset{G}{\iota}=\iota\circ g.$

Now, using the property: `inverse of composition equals composition of inverse
into reversed order', we finally get
\[
\underset{G}{\iota}^{-1}=(\iota\circ g)^{-1}=g^{-1}\circ\iota^{-1}.
\]
\end{proof}

\section{Conclusions}

We investigated the relationship between metric tensors and metric extensors
associated to a $n$-dimensional real vector space, and translated some well
known results of tensor theory using extensors. The results obtained,
specially eq.(\ref{3.2.1}) that relates the classical determinant of the
matrix whose entries are the components of a $2$-tensor in a given vector
basis with the determinant of the corresponding extensor, is important for
many calculations which will appear in the next papers reporting ours studies
on the theory of multivector functions, and some problems of differential
geometry and theoretical physics. We emphasize moreover that even in
elementary linear algebra it is an advantage to use metric extensors instead
of metric tensors because a metric extensor has an inverse\footnote{Of course,
a metric tensor has no inverse, see \cite{1}.} which is itself a metric
extensor. Also, with the concept of metric extensor sophisticated calculations
can be done without the introduction of matrix representations.\emph{\vspace
{0.1in}}

\textbf{Acknowledgement}: V. V. Fern\'{a}ndez is grateful to FAPESP for a
posdoctoral fellowship. W. A. Rodrigues Jr. is grateful to CNPq for a senior
research fellowship (contract 251560/82-8) and to the Department of
Mathematics of the University of Liverpool for the hospitality. Authors are
also grateful to Drs. P. Lounesto, I. Porteous and J. Vaz Jr. for their
interest in our research and useful discussions.

\end{document}